\documentclass[twocolumn,showpacs,aps,prl,floatfix,superscriptaddress]{revtex4}
\expandafter\let\csname equation*\endcsname\relax
\expandafter\let\csname endequation*\endcsname\relax

\usepackage{graphicx,enumerate,bm,amsbsy}
\usepackage{amsmath,amssymb,wasysym,graphicx,capt-of,ifthen,calc}
\usepackage{amsfonts}
\usepackage{bm}
\usepackage{dcolumn}
\usepackage{enumitem}
\usepackage{subfigure}
\usepackage{verbatim} 
\usepackage{color}
\usepackage{float}
\usepackage{url}
\usepackage{hyperref}

\newcommand{\gl}{\mathrel{\raise0.6ex\hbox{$>$\kern-.75em\lower1ex\hbox{$<$}}}}

\begin{document}
\title{Highly Dispersed Networks}
\author{Alan Gabel}
\affiliation{Center for Polymer Studies and Department of Physics, Boston University, Boston, MA 02215, USA}
\author{P. L. Krapivsky}
\affiliation{Department of Physics, Boston University, Boston, MA 02215, USA}
\author{S. Redner}
\affiliation{Center for Polymer Studies and Department of Physics, Boston University, Boston, MA 02215, USA}

\begin{abstract}
  We introduce a new class of networks that grow by \emph{enhanced redirection}.
  Nodes are introduced sequentially, and each either attaches to a randomly
  chosen target node with probability $1-r$ or to the ancestor of the target
  with probability $r$, where $r$ an increasing function of the degree of the
  ancestor.  This mechanism leads to highly-dispersed networks with unusual
  properties: (i) existence of multiple macrohubs---nodes whose degree is a
  finite fraction of the total number of network nodes $N$, (ii) lack of self
  averaging, and (iii) anomalous scaling, in which $N_k$, the number of nodes
  of degree $k$ scales as $N_k\sim N^{\nu-1}/k^{\nu}$, with $1<\nu<2$.
\end{abstract}

\maketitle

Many of the current models for complex networks involve growth mechanisms
that are based on \emph{global} knowledge of the network.  For example, in
preferential attachment~\cite{BA99,DM03,N10}, a new node attaches to an
existing node of the network at a rate that is a monotonically increasing
function of the degree of this target node.  To implement this rule
faithfully, one needs to know the degree of every node in the network, and it
is impractical to maintain such detailed knowledge of the network.

A counterpoint to global growth rules is provided by a class of models that
require only \emph{local} knowledge of the network, including, for example,
spatial locality\cite{FKP02,CBMR04,B11} and node similarity~\cite{PKSBK12}.
An appealing model of this genre is
redirection~\cite{KR01,V03,RA04,KR05,LA07,BK10}.  Here, each newly-introduced
node chooses a target node at random and attaches to this target and/or to
one or more of its ancestors.  If redirection occurs only to an immediate
ancestor with a fixed probability, the resulting growth rule corresponds
exactly to shifted linear preferential attachment~\cite{KR01}.  Two important
features of this redirection mechanism are: (i) it precisely mimics global
growth rules, such as preferential attachment, and (ii) efficiency, as the
addition of each node requires just two computer instructions, so that the
time needed to simulate a network of $N$ nodes scales linearly with $N$.

\begin{widetext}

\begin{figure}[ht]
\begin{center}
\subfigure[]{\includegraphics[width=0.3\textwidth]{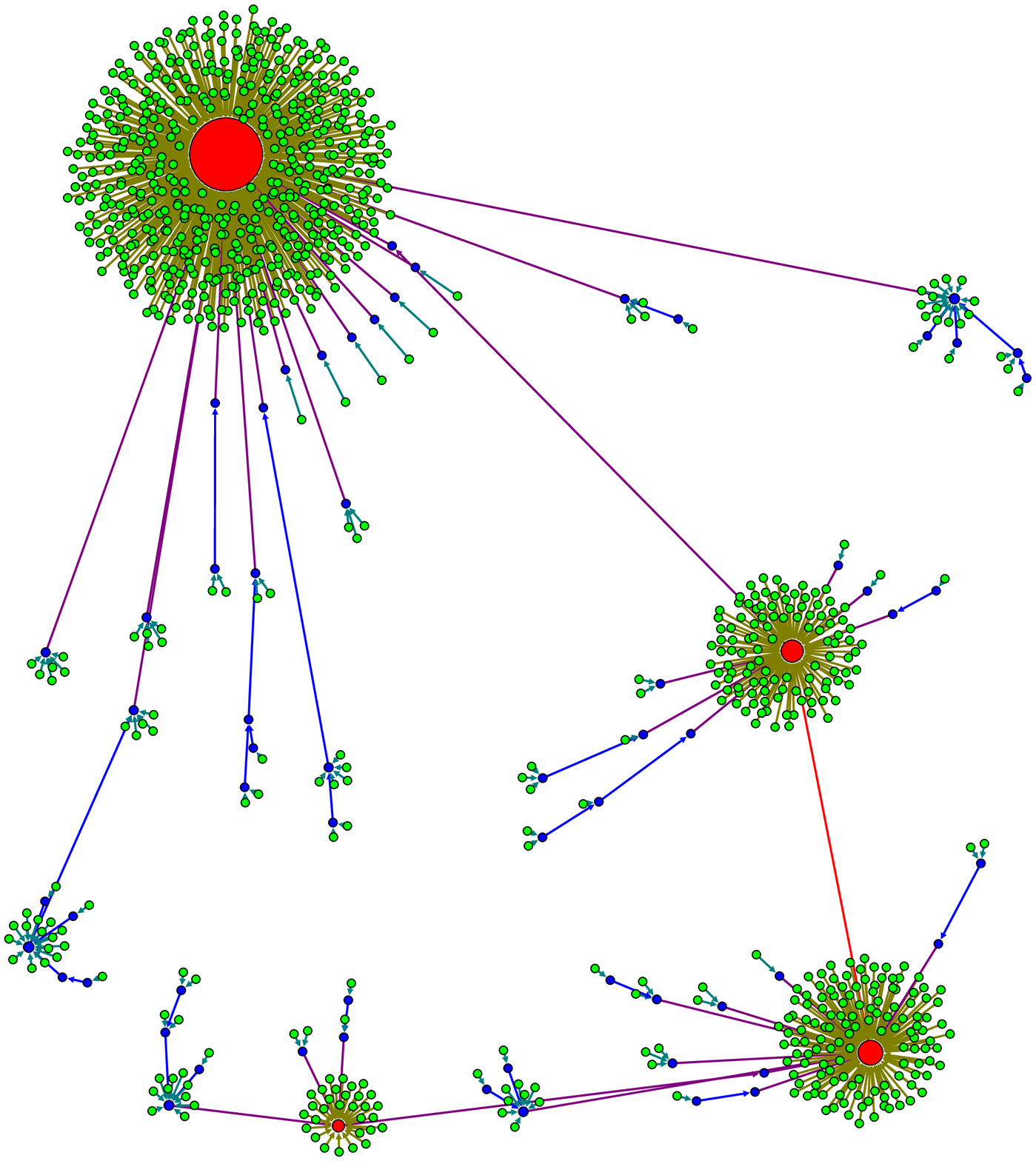}} \quad
\subfigure[]{\includegraphics[width=0.3\textwidth]{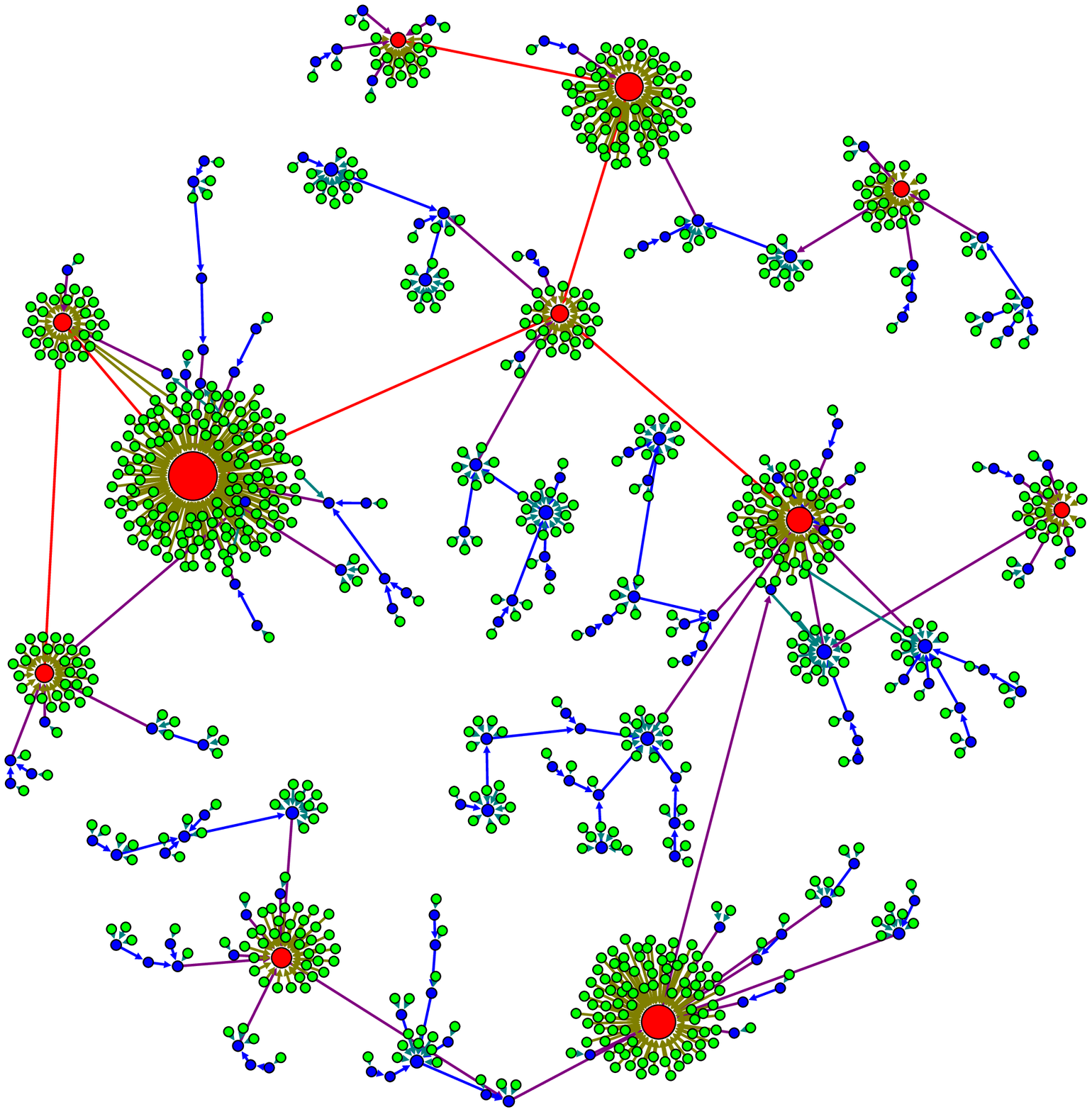}} \quad
\subfigure[]{\includegraphics[width=0.3\textwidth]{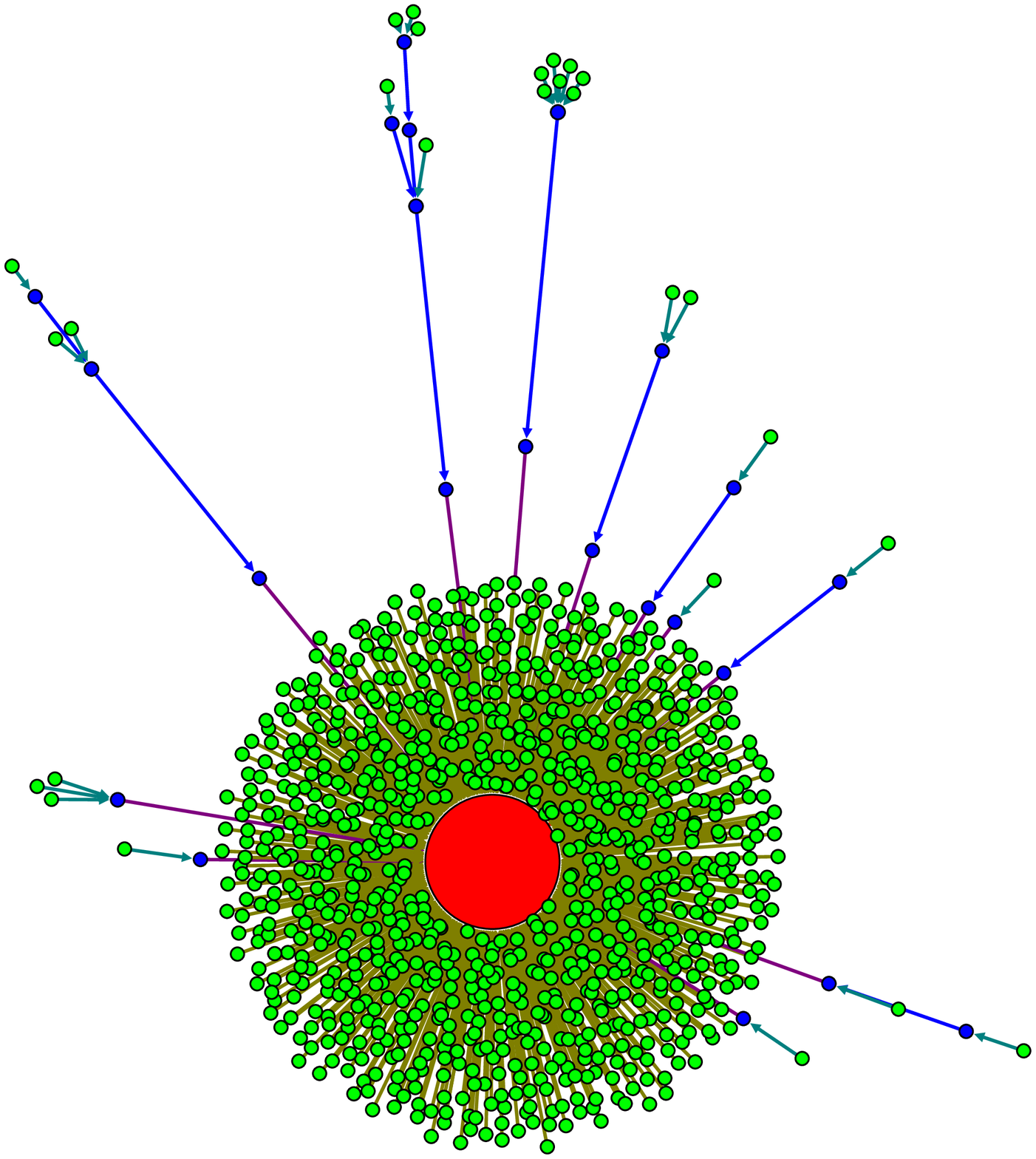}}
\caption{Enhanced redirection networks of $N=10^3$ nodes for
  $\lambda=\frac{3}{4}$ starting from the same initial state.  (a) Maximum
  degree $k_{\rm max}=548$, $\mathcal{C}=66$ core (degree $\geq 2$) nodes,
  and maximum diameter $D_{\rm max}=10$.  (b) $k_{\rm max}=\mathcal{C}=154$,
  $D_{\rm max}=12$ (the smallest $k_{\rm max}$ out of $10^3$ realizations).
  (c) $k_{\rm max}=963$, with $\mathcal{C}=23$ and $D_{\rm max}=6$ (the
  largest $k_{\rm max}$ out of $10^3$ realizations).  Green: nodes of degree
  1, blue: degrees 2--20, red: degree $>20$.  The link color is the average
  of the endpoint nodes.}
  \label{typical}
\end{center}
\end{figure}

\end{widetext}

The utility of redirection as an efficient way to mimic linear preferential
attachment motivates us to exploit slightly more, but still local,
information around the target node.  Specifically, we consider a redirection
probability $r(a,b)$ that depends on the degrees of the target and ancestor
nodes, $a$ and $b$ respectively.  In \emph{hindered redirection}, $r(a,b)$ is
a decreasing function of the ancestor degree $b$~\cite{GR13}, a rule that
leads to sub-linear preferential attachment network growth.  In this work, we
investigate the complementary situation of \emph{enhanced redirection}, for
which the redirection probability $r$ is an increasing function of the
ancestor degree $b$ with $r\to 1$ as $b\to\infty$.  This seemingly-innocuous
redirection rule gives rise to networks with several intriguing and
practically relevant properties (Fig.~\ref{typical}):
\begin{itemize}
\item Multiple \emph{macrohubs}---whose degrees are a finite fraction of
  $N$---arise.  
\item Lack of self averaging.  Different network realizations are visually
  diverse when the growth process starts from the same initial condition, in
  contrast to preferential attachment~\cite{KR02f}.
\item Non-extensivity.  The number of nodes of degree $k$, $N_k$, scales as
  $N_k\sim N^{\nu-1}k^{-\nu}$, with $1<\nu<2$, again in contrast to
  preferential attachment, where $N_k\sim Nk^{-\nu}$ with $\nu>2$.
\end{itemize}

Several aspects of these novel attributes bear emphasis.  While macrohubs
also occur in superlinear preferential attachment~\cite{KR01,KRL00,KK08} and
in the fitness model~\cite{BB01,KR02}, these examples give a single
macrohub.  In contrast, enhanced redirection networks are highly disperse,
with interconnected hub-and-spoke structures that are reminiscent of airline
route networks~\cite{N10,BO99,CS03,H04,GMTA05}.  Regarding non-extensive scaling,
a degree exponent in the range $1<\nu<2$ has been observed in numerous
networks \cite{KBM13}.  Taken together with extensivity so that $N_k\sim
Nk^{-\nu}$, the range $1<\nu<2$ is mathematically inconsistent.  Namely, for
sparse networks the average degree is finite, while $\langle k\rangle =
N^{-1}\sum_{k=1}^N kN_k$ diverges as $N^{2-\nu}$. The simplest resolution of
this paradox is to posit
\begin{equation}
\label{NkN}
N_k\sim N^{\nu-1}k^{-\nu}
\end{equation}
for $k\geq 2$. The number of nodes of degree 1 (leaves) $N_1$ must still grow
linearly with $N$ so that the sum rule $\sum_{k=1}^N N_k=N$ is obeyed.  More
precisely, the scaling with system size is
\begin{equation}
\label{scaling}
N-N_1 = \mathcal{O}(N^{\nu-1}), \quad  N_k = \mathcal{O}(N^{\nu-1}) ~~k\geq 2
\end{equation}

In our modeling, links are directed and each node has out-degree equal to 1,
and thus a unique ancestor.  This growth rule produces tree networks; closed
loops can be generated by allowing each new node to connected to the network
in multiple ways~\cite{RA04}.  For convenience, we choose the initial
condition of a single root node of degree 2 that links to itself.  The root
is thus both its own ancestor and its own child.  Nodes are introduced one by
one.  Each first picks a random target node (of degree $a$), and then:
\begin{enumerate}
\itemsep -0.3ex
\item[(a)] either the new node attaches to the target with probability
  $1-r(a,b)$;
\item[(b)] or the new node attaches to the ancestor (with degree $b$) of the
  target with probability $r(a,b)$.
\end{enumerate}

The unexpected connection between constant redirection probability and
shifted linear preferential attachment~\cite{KR01,RA04} arises because the
number of ways to redirect to an ancestor node is proportional to the number
of its descendants and thus to its degree.  For enhanced redirection, two
natural (but by no means unique) choices for the redirection probability are:
\begin{equation}
\label{r}
r(a,b)=1-b^{-\lambda},\qquad r(a,b)=\frac{a^\lambda}{a^\lambda+b^\lambda},\qquad \lambda>0\,.
\end{equation}
Our results are robust with respect to the form of the redirection
probability, as long as $r(a,b)\to 1$ as $b\to\infty$; we primarily focus on
the first model.  


We now present analytical and numerical evidence for the emergence of
macrohubs, the lack of self-averaging, and non-extensivity, as embodied by
Eq.~\eqref{scaling}.

\begin{figure}[ht]
\begin{center}
\subfigure[]{\includegraphics[width=0.235\textwidth]{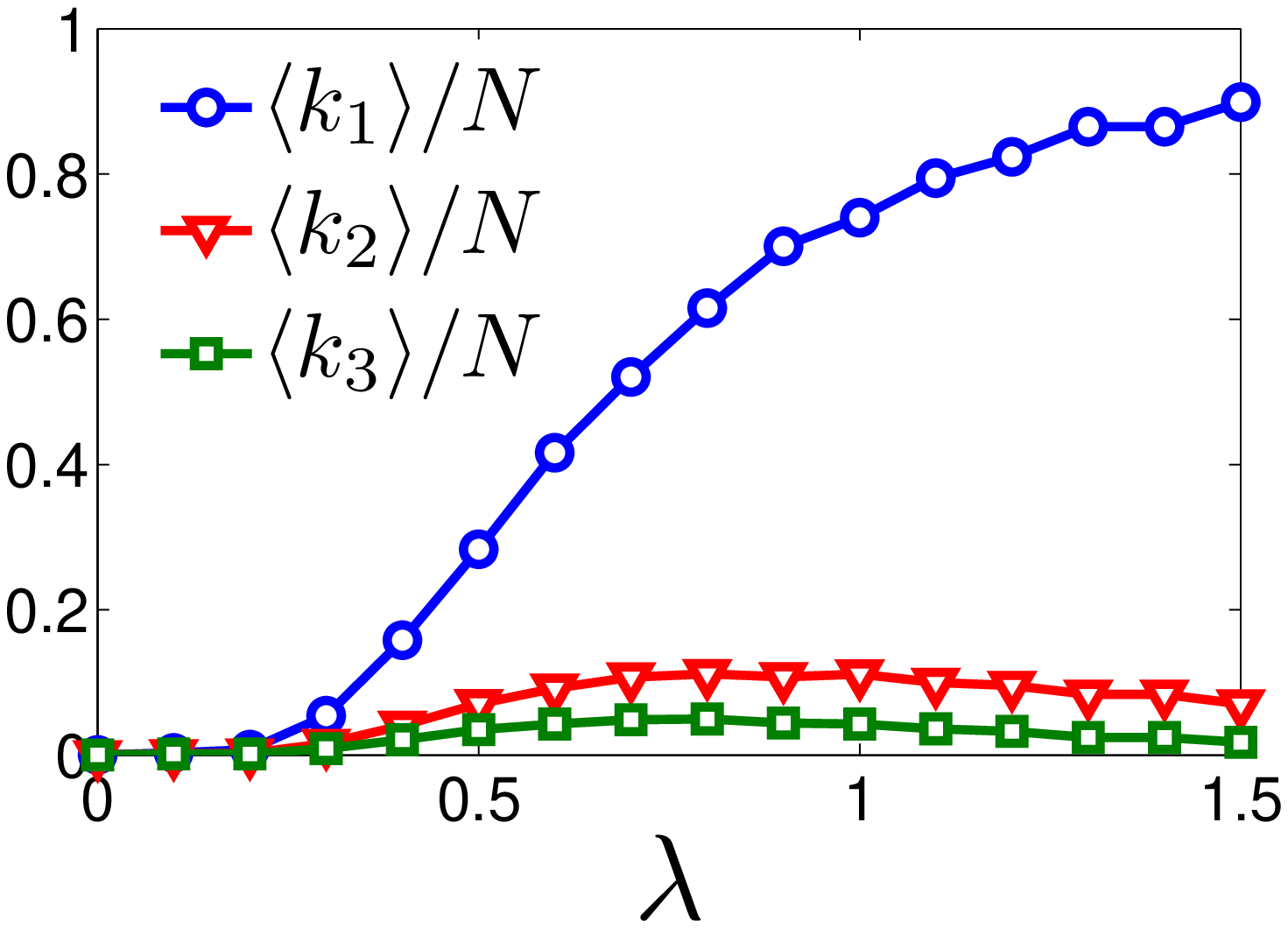}}
\subfigure[]{\raisebox{0.15 em}{\includegraphics[width=0.23\textwidth]{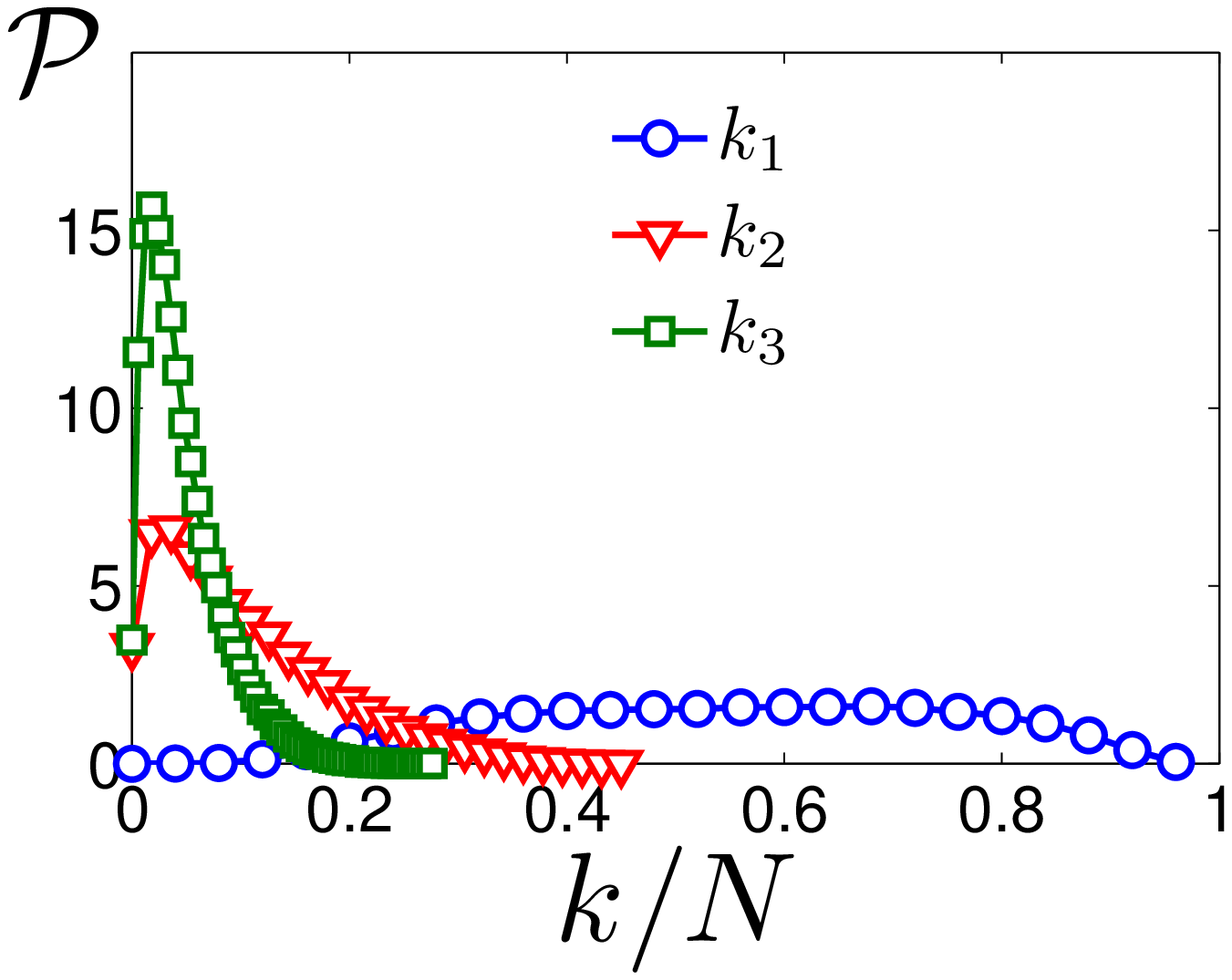}}}
\caption{(a) Average value of the three largest degrees (divided by $N$) as a
  function of $\lambda$.  Each data point corresponds to $10^3$
  realizations. (b) Probability densities of these three largest degrees for
  $\lambda=\frac{3}{4}$. }
  \label{kmPlot}
\end{center}
\end{figure}

{\it Macrohubs:} Macrohubs inevitably arise in all network realizations.
Figure~\ref{kmPlot}(a) shows that the average largest, $2^{\rm nd}$-largest,
and $3^{\rm rd}$-largest degrees are all macroscopic. These degrees, as well
as the degrees of smaller hubs, are broadly distributed
(Fig.~\ref{kmPlot}(b)).  We estimate the maximum degree $k_{\rm max}$ by the
extremal criterion: $\int_{k_{\rm max}}^\infty N_k\, dk \sim 1$.  For
$N_k\sim N^{\nu-1}k^{-\nu}$ and $1<\nu<2$, this criterion gives $k_{\rm
  max}\sim N$.  In contrast, for linear preferential attachment
with $N_k\sim Nk^{-3}$,  $k_{\rm max}\sim N^{1/2}$.

\begin{figure}[ht]
\begin{center}
\includegraphics[width=0.3\textwidth]{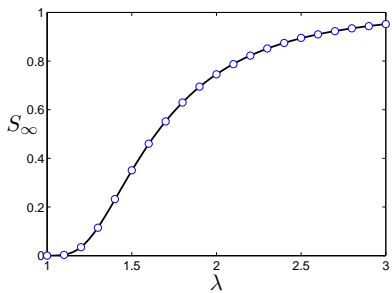}
\caption{Probability for a star graph, $S_\infty$, versus $\lambda$.  Data
  points are based on $10^4$ realizations for each $\lambda$.  The curve is
  the numerical evaluation of the product in~\eqref{S_max}. }
  \label{star}
\end{center}
\end{figure}

The dominant role of macrohubs can be appreciated by computing the
probability that the node with the highest degree attaches to every other
node of the network, thereby making a star.  Suppose that the network has $N$
nodes and still remains a star.  For the initial condition of a single node
with a self loop, this star graph contains $N-1$ leaves and the hub has
degree $N+1$.  The probability $S_{N}$ to build such a graph is
\begin{align}
\label{S_max}
S_{N}(\lambda) = \prod_{n=1}^{N-1}\left\{\frac{1}{n}+\frac{n-1}{n}\left[1 -
    (n+1)^{-\lambda} \right]\right\}.
\end{align}
The factor $\frac{1}{n}$ accounts for the new node attaching to the root in a
network of $n$ nodes, while the second term accounts for first choosing a
leaf and then redirecting to the root.  The asymptotic behavior of
\eqref{S_max} is:
\begin{equation}
\label{S_max_asymp}
S_{N}(\lambda) \to 
\begin{cases}
S_\infty(\lambda)     & \lambda>1\\
A/N                     & \lambda=1\\
 \exp\!\left(-\frac{N^{1-\lambda}}{1-\lambda}\right)    &0<\lambda<1\\
\frac{1}{(N-1)!}    & \lambda=0,
\end{cases}
\end{equation}
where $0<S_\infty(\lambda)<1$, and $A=\pi^{-1}\sinh\pi\approx 3.676$.  Thus a
star graph occurs with positive probability when $\lambda>1$ and
$S_\infty(\lambda)$ quickly approaches 1 as $\lambda$ increases
(Fig.~\ref{star}).  This makes obvious the emergence of hubs for $\lambda>1$.
A more detailed analysis is required for $0<\lambda\leq 1$ that also confirms
the inevitability of macrohubs.  We find that the probability
$P(k_\text{max})$ that the maximal degree is $k_\text{max}$ has the scaling
form \cite{GKR}
\begin{equation}
\label{scaling:max}
P(k_\text{max})\to \frac{1}{N}\,\mathcal{P}(x), \quad x = \frac{k_\text{max}}{N}
\end{equation}
with $\mathcal{P}(1)=A$ for $\lambda=1$, while for $0<\lambda<1$ the scaling
function vanishes as $\ln \mathcal{P}(x)\sim - (1-x)^{\lambda -1}$ when $x\to
1$.

{\it Non Self Averaging:} Enhanced redirection networks display huge sample
to sample fluctuations (Fig.~\ref{typical}), as exemplified by
\eqref{scaling:max}.  Another manifestation of these fluctuations is provided
by the distributions for the fraction of nodes of fixed degree $k$,
$P(N_k/N)$.  For preferential attachment networks, this distribution becomes
progressively sharper as $N$ increases~\cite{KR02}, as long as the degree is
not close to its maximal value.  Thus the average fractions of nodes of a
given degree constitute the set of variables that characterizes the degree
distribution; only the nodes with the highest degrees fail to self
average~\cite{KR02_b}.

\begin{figure}[ht]
\begin{center}
\subfigure[]{\raisebox{1.15 em}{\includegraphics[width=0.235\textwidth]{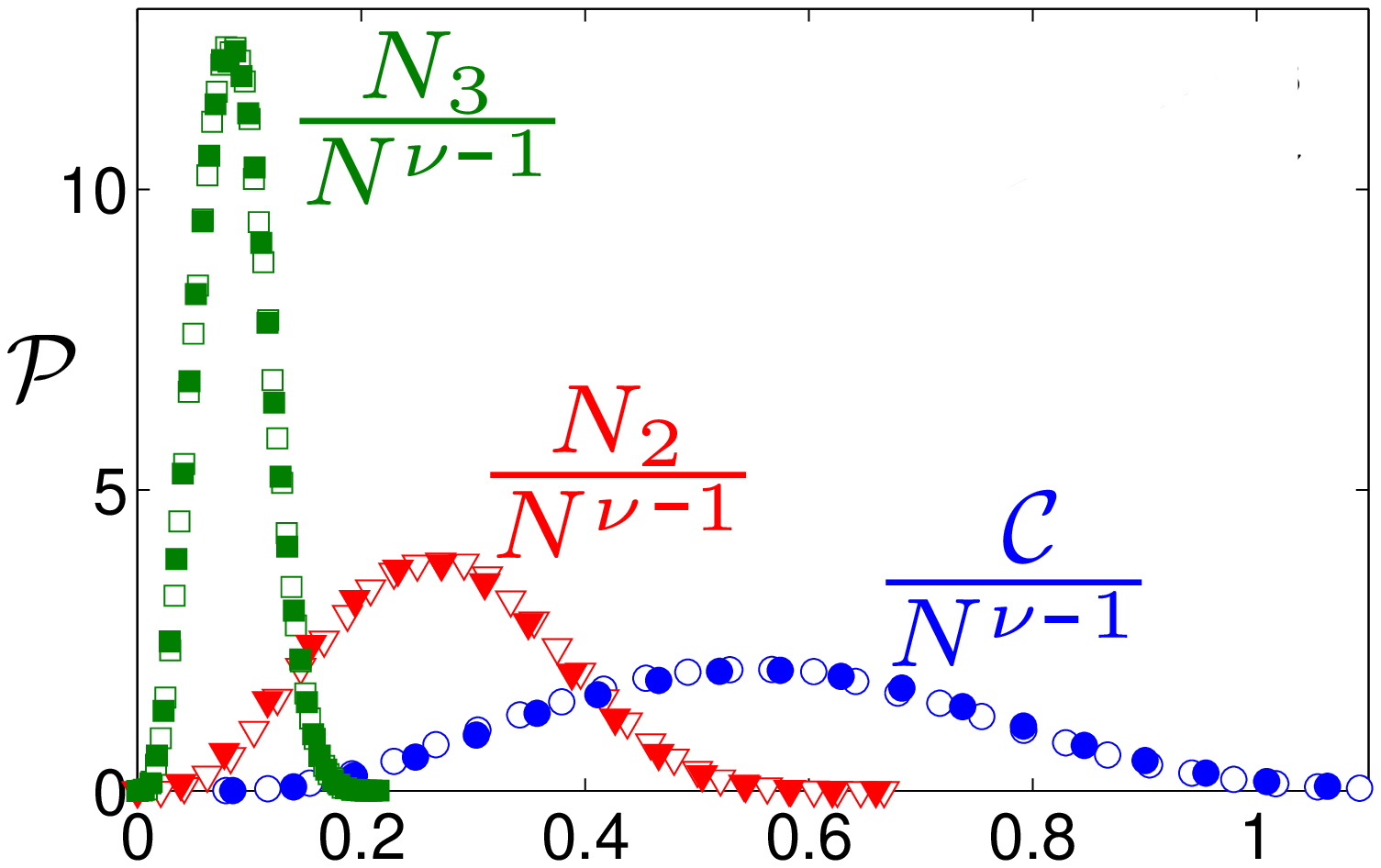}}}
\subfigure[]{\includegraphics[width=0.235\textwidth]{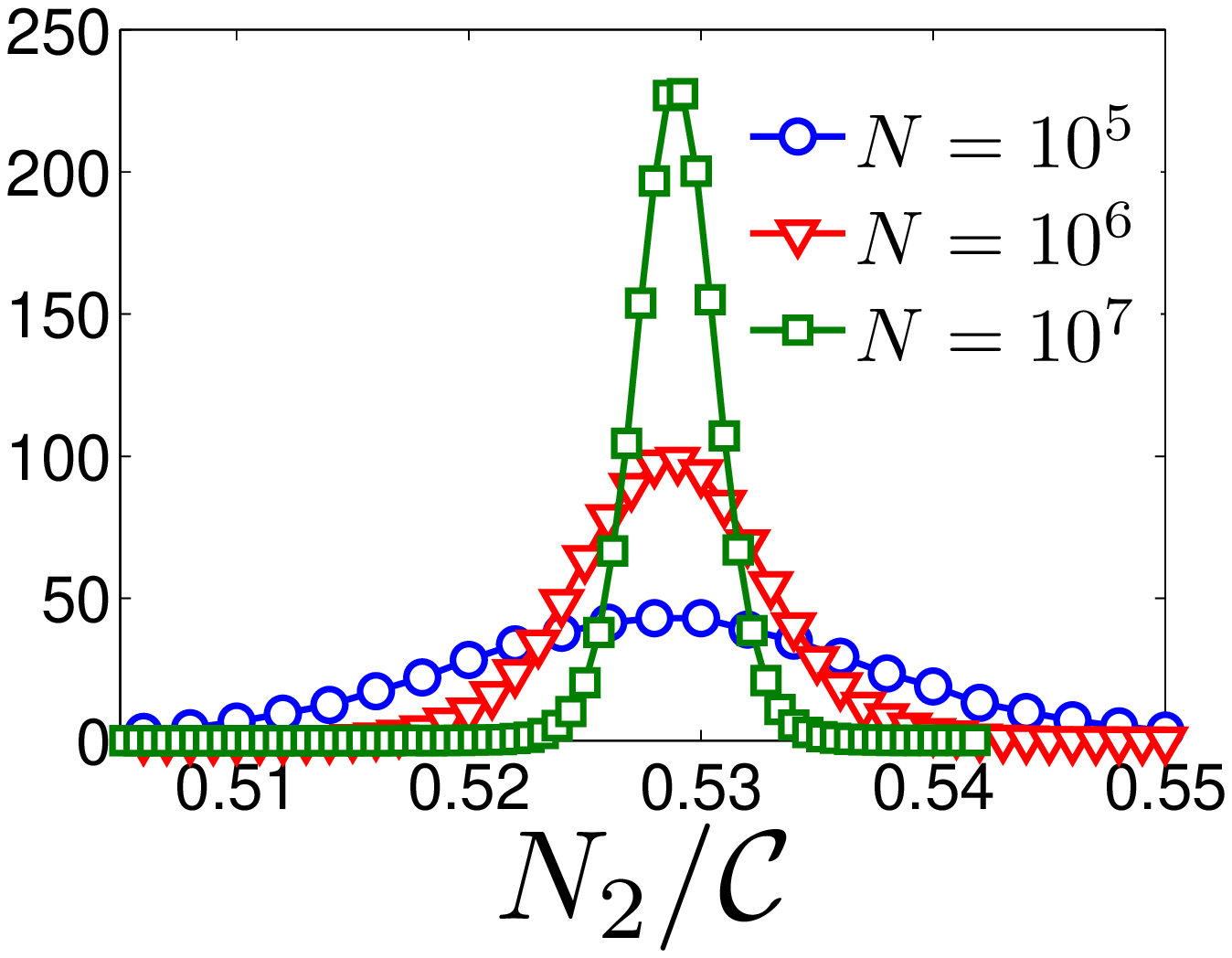}}
\caption{Probability densities for enhanced redirection for: (a)
  $\mathcal{C}/N^{\nu\!-\!1}$, $N_2/N^{\nu\!-\!1}$, and $N_3/N^{\nu\!-\!1}$
  for $N=10^6$ (open) and $N=10^7$ nodes (closed symbols) and (b)
  $N_2/\mathcal{C}$.  Data are based on $10^5$ realizations with
  $\lambda=\frac{3}{4}$ and $\nu=1.73$.}
  \label{NkDist}
\end{center}
\end{figure}

In contrast, for enhanced redirection networks, essentially all geometrical
features are non self-averaging.  Figure~\ref{NkDist} shows the distributions
of $\mathcal{C}/N^{\nu-1}$, $N_2/N^{\nu-1}$, $N_3/N^{\nu-1}$, etc., which do
not sharpen as $N$ increases.  Here $\mathcal{C}\equiv N-N_1$ is the number
of non-leaf (``core'') nodes.  Since $\mathcal{C}$ and $N_k$ for $k\geq 2$
all scale as $N^{\nu-1}$ (Eq.~\eqref{scaling}), appropriately scaled
distributions of these quantities would progressively sharpen as $N$
increases if self averaging holds.  


Surprisingly, the ratios $N_k/\mathcal{C}$ are self-averaging for $k\ge 2$, as
the distributions $N_k/\mathcal{C}$ do sharpen as $N$ increases
(Fig.~\ref{NkDist}).  The self-averaging of these ratios suggests that
although the overall number of core nodes $\mathcal{C}$ varies widely between
realizations, the degree distributions \emph{given} a value of
$\mathcal{C}$ are statistically the same.

We can understand the lack of self averaging in enhanced redirection networks
in a heuristic way.  Once a set of macrohubs emerges (with degrees $k_1$,
$k_2$, $k_3$, $\dots$) the probability of attaching to a macrohub of degree
$k_i$ asymptotically approaches $k_i$.  This preferential attachment to
macrohubs is precisely the same prescription for a multistate P\'olya urn
process for filling an urn with balls of several colors~\cite{EP23,urn}, for
which it is known that the long-time distribution of the number of balls of a
given color is a non self-averaging quantity.

\emph{Degree Distribution:} Since the degree distribution itself is
non self averaging, we focus on the average over all realizations, $\langle
N_k\rangle$.  To avoid notational clutter we write $N_k$ for the average
$\langle N_k\rangle$.  Each time a new node is introduced, the degree evolves
according to
\begin{align}
\label{master1}
\frac{dN_k}{dN}&=\frac{(1-f_{k-1})N_{k-1}-(1-f_k)N_k}{N} \nonumber \\
&\quad +\frac{(k-2)t_{k-1}N_{k-1}-(k-1)t_kN_k}{N}+\delta_{k,1}\,.
\end{align} 
Here $f_k$ and $t_k$ are defined as the respective probabilities that an
incoming link is redirected \emph{from} a node of degree $k$, and \emph{to} a
node of degree $k$.  The terms involving the factor $1\!-\!f_j$ in
\eqref{master1} thus account for events where the incoming link attaches
directly to the target, while the terms involving the factor $t_j$ account
for redirection to the ancestor.  The term $\delta_{k,1}$ accounts for the
new node of degree $1$.  Defining $\alpha_k=(k\!-\!1)t_k+1-f_k$,
Eq.~\eqref{master1} can be written in the canonical form
\begin{align}
\label{master2}
\frac{dN_k}{dN}=\frac{\alpha_{k-1}N_{k-1}-\alpha_kN_k}{N}+\delta_{k,1}\,.
\end{align} 

\begin{figure}[ht]
\begin{center}
\subfigure[]{\includegraphics[width=0.235\textwidth]{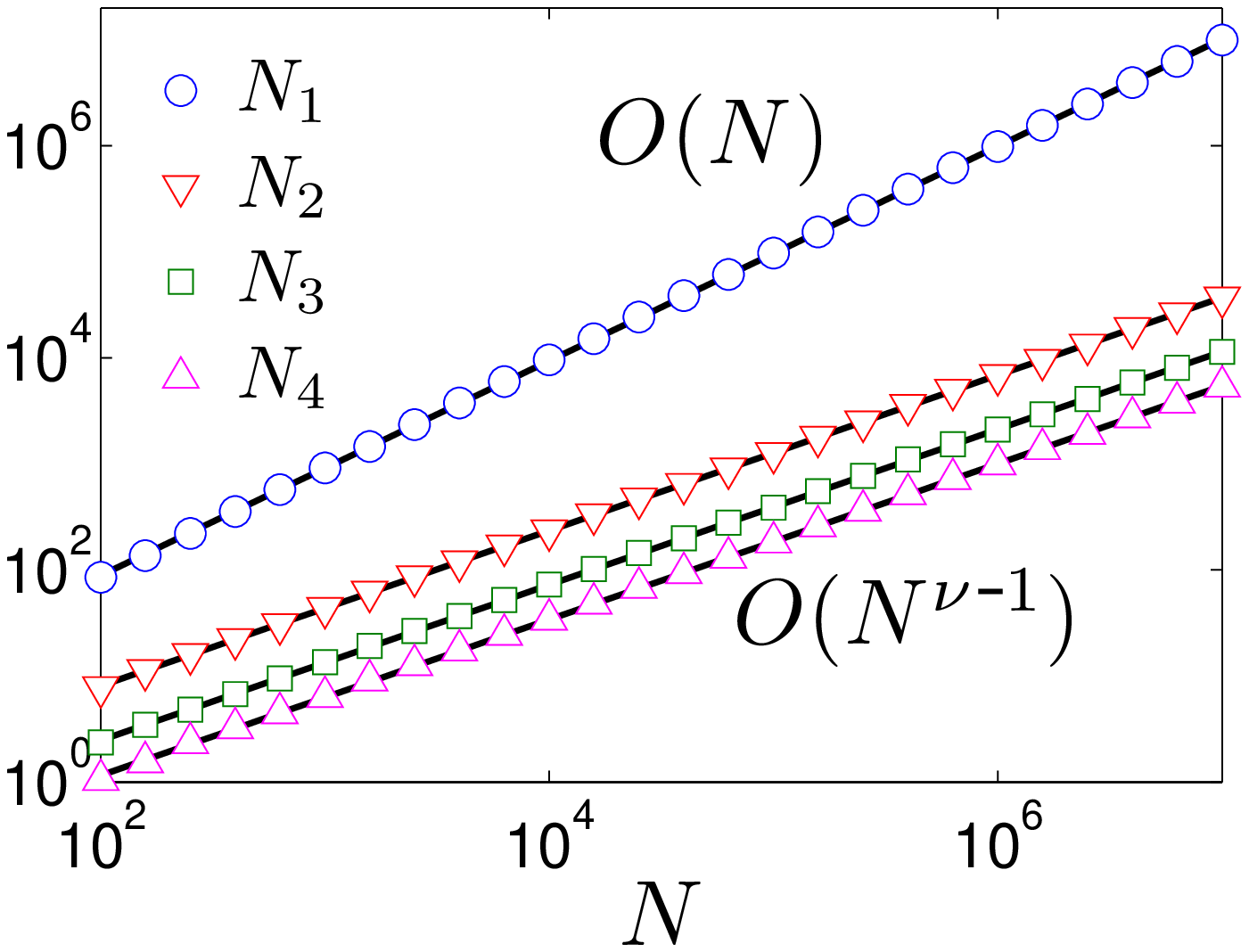}}
\subfigure[]{\includegraphics[width=0.235\textwidth]{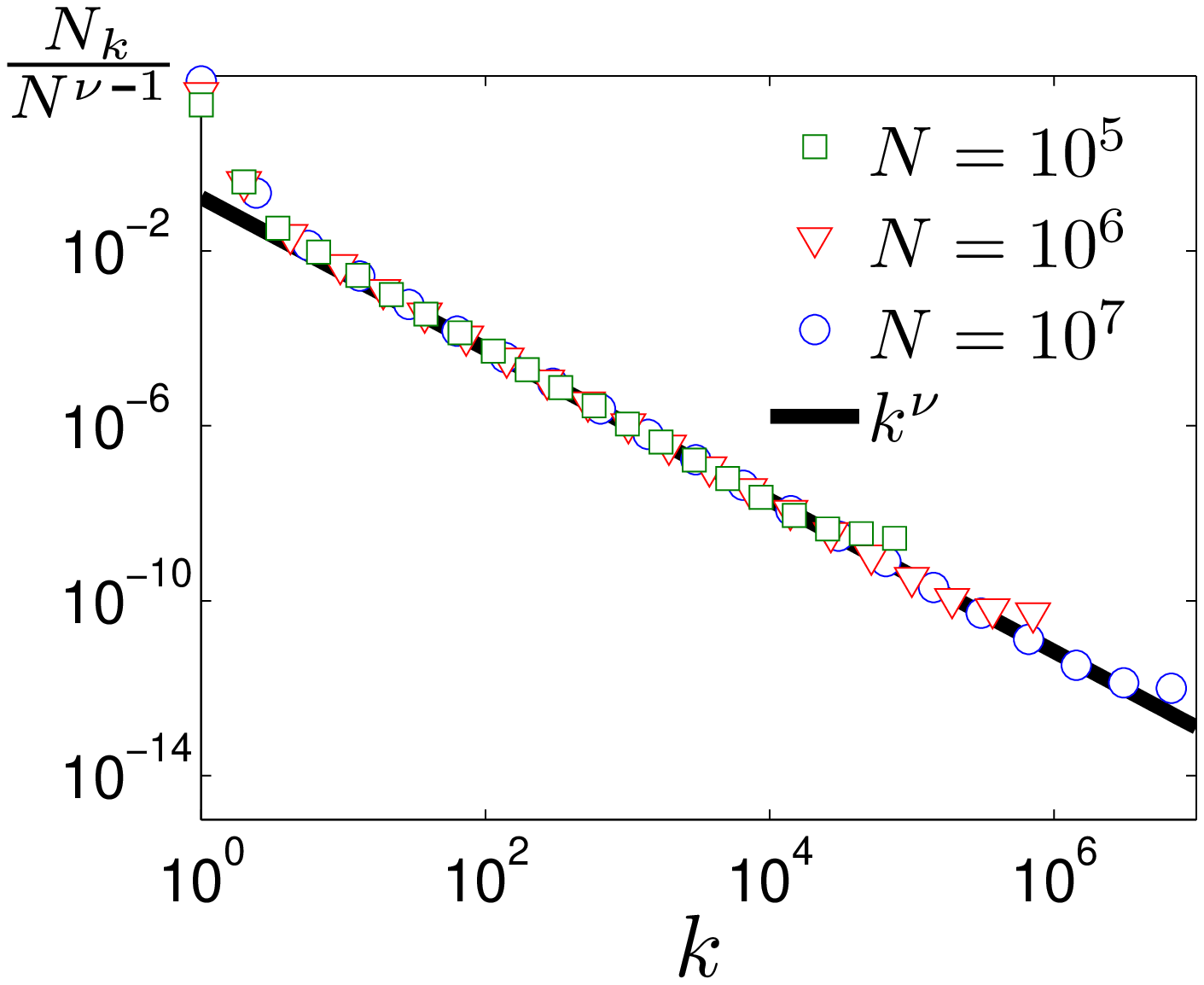}}
\caption{(a) $N_k$ versus $N$ and (b) $N_k/N^{\nu-1}$ versus $k$ for enhanced
  redirection with $\lambda=\frac{3}{4}$ and $\nu=1.73$ (determined
  numerically; Fig.~\ref{nu}).  Data are based on $10^4$ realizations, with
  equally-spaced bins on a logarithmic scale in (b).  The lines in (a) show
  the prediction of Eq.~\eqref{ansatz}, while the line in (b) shows the $k$
  dependence from the numerical solution of \eqref{ck}.  }
  \label{NkVsN}
\end{center}
\end{figure}

We now use the empirically-observed scaling \eqref{scaling}, as illustrated
in Fig.~\ref{NkVsN}, to deduce the algebraic decay \eqref{NkN}. We rewrite
\eqref{scaling}  more precisely as
\begin{equation}
\label{ansatz}
N-N_1\simeq c_1N^{\nu-1},\qquad N_k\simeq c_kN^{\nu-1} \quad k\ge 2\,,
\end{equation}
and substitute it into into the evolution equations \eqref{master1}.
Straightforward calculation gives the product solution
\begin{equation}
\label{ck}
c_k=c_1\,\frac{\nu-1}{\alpha_k}\prod_{j=2}^k \left( \frac{\alpha_j}{\alpha_j+\nu-1} \right).
\end{equation}
We now need the analytic form for $\alpha_k$, which requires the
probabilities $f_k$ and $t_k$.  The latter are given by
\begin{equation*}
f_k=\sum_{b\ge 1} \frac{r(k,b)N(k,b)}{N_k}\,,\quad 
t_k=\sum_{a\ge 1}\frac{r(a,k)N(a,k)}{(k-1)N_k}\,,
\end{equation*}
where $N(a,b)$ is the number of nodes of degree $a$ that have an ancestor of
degree $b$.  Thus $f_k$ is the probability of redirecting \emph{from} a node
of degree $k$, averaged over all such target nodes, and $t_k$ is the
probability of redirecting \emph{to} a node of degree $k$, averaged over all
the $(k-1)N_k$ children of nodes of degree $k$.

\begin{figure}[ht]
\begin{center}
\includegraphics[width=0.3\textwidth]{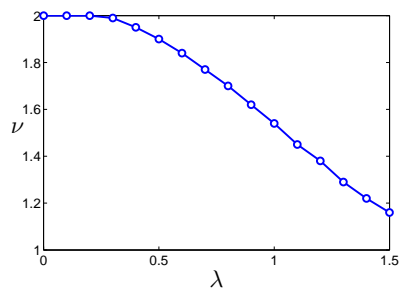}
\caption{Degree distribution exponent $\nu$ versus $\lambda$.  Each data
  point is determined from fits of $N_k$ versus $N$, as in
  Fig.~\ref{NkVsN}(a). }
  \label{nu}
\end{center}
\end{figure}

For redirection probability $r(a,b)=1-b^{-\lambda}$, the probabilities $f_k$
and $t_k$ reduce to
\begin{align*}
f_k=\sum_{b\ge 1} \frac{(1-b^{-\lambda})N(k,b)}{N_k}\equiv 1 -\langle
b^{-\lambda}\rangle\,,\nonumber \\
t_k=\sum_{a\ge 1}\frac{(1-k^{-\lambda})N(a,k)}{(k-1)N_k}=1-k^{-\lambda}\,,
\end{align*}
leading to $\alpha_k=k-k^{1\!-\!\lambda}+k^{-\lambda}-f_k\to k$ in the
large-$k$ limit.  Using $\alpha_k \sim k$ in the product solution \eqref{ck}
gives the asymptotic behavior
\begin{align}
\label{largeK}
  c_k\sim c_1\,\frac{\nu\!-\!1}{k}\,\prod_{j=2}^k \left( \frac{j}{j+\nu-1}  \right)
  &\sim k^{-\nu}\,.
\end{align}
Thus the degree distribution exhibits anomalous scaling, $N_k\sim
N^{\nu-1}k^{-\nu}$, with $1<\nu<2$.  Numerical simulations show that the
exponent $\nu$ is a decreasing function of $\lambda$ and that $\nu\to 1$ as
$\lambda\to 2$ (Fig.~\ref{nu}).  There is clear evidence of a transition at
$\lambda=2$; for larger $\lambda$, nodes of degree no longer appear.  Thus
the network consists of a collection ``hairballs''---star graphs that
are connected to each other by single links.

To conclude, enhanced redirection is a simple and appealing mechanism that
produces networks with several anomalous features that are observed in real
networks.  Among them are the existence of multiple macroscopic hubs, as
arises in the airline route network~\cite{N10,BO99,CS03,H04,GMTA05}.  Thus
our networks are highly disperse and consist of a set of loosely-connected
macrohubs (Fig.~\ref{typical}).  Also intriguing is the anomalous scaling of
the degree distribution, in which the number of nodes of degree $k$ decays
more slowly than $k^{-2}$.  Such a decay is mathematically possible in sparse
networks only if the number of nodes of any degree scales sublinearly with
$N$.  Enhanced redirection may thus provide the mechanism that underlies the
wide range of networks~\cite{KBM13} whose degree distributions apparently
decay more slowly than $k^{-2}$.

\bigskip\noindent We gratefully acknowledge financial support from grant
\#FA9550-12-1-0391 from the U.S.  Air Force Office of Scientific Research
(AFOSR) and the Defense Advanced Research Projects Agency (DARPA).

\end{document}